\documentstyle[preprint,balanced,aps,prl,epsfig]{revtex}
\begin{document}
\title{Noise-induced  memory in extended excitable systems }
\author{Dante R. Chialvo , Guillermo A. Cecchi ,  Marcelo O. Magnasco}
\address{{\sl $^{\dagger}$Center for Studies in Physics and Biology,
The Rockefeller University, 1230 York Avenue, New York NY10021, USA}}
\date{\today}
\maketitle
\begin{abstract}
We describe a form of memory exhibited by extended excitable systems
driven by stochastic fluctuations.
Under such conditions, the system self-organizes into a state
characterized by power-law correlations thus retaining
long-term memory of previous states.
The exponents are robust and model-independent.
We discuss novel implications of these results for the functioning of
cortical neurons as well as for networks of neurons.
\end{abstract}

Neurons receive thousands of perturbations affecting the transmembrane voltage
at various points of the synaptic membrane. Recent experimental evidence has
shown active nonlinearities \cite{active} at the dendrites of cortical neurons,
and thus implying that a model representing these neurons must have many
nonlinear spatial degrees of freedom.

What are the dynamical consequences of these distributed nonlinearities for the
neurons function? The answer is not inmediately certain. The prevailing view
has been, since Lapicque in 1907 \cite{lapique}, that all input regions (i.e.,
dendrites) were linear, and thus neurons were represented as  a single
compartment. In this view incoming excitations are linearly integrated and 
whenever the resulting value exceeds a predefined threshold an output (action
potential) is generated. Thus, the neuron is considered to has a single non-linear
degree of freedom (i.e., the spatial region where the thresholding dynamics
takes place).

This Letter describes a robust form of noise-induced memory which appears
naturally as a direct consequence of including distributed nonlinearities in
the formulation of a neuron's input region. Besides having relevance at the
neural level, its touches other areas of biology where excitable models have
been used, as is the case for models of forest-fire propagation, spreading of
epidemics and noise-induced waves.\cite{waves} From the outset, it needs to be
noted that the phenomena to be described do not depend of the type of
excitable model one uses.

To show the essence of the main point we  adapt the Greenberg-Hastings cellular
automata model\cite{model} of excitable media\cite{othermodels}. For the
purpose of this report let restrict ourselves to the case  of a one-dimensional
lattice  of coupled identical compartments ($n=1,...,N$), with open boundary
conditions. Thus, each spatial location is assigned a  discrete state $S^t_n$
which can be one of  three: Quiescent, Excited or Refractory, with the dynamics
determined by the transition rules: E $\rightarrow$ R (always), R $\rightarrow$
Q (always), Q $\rightarrow$ E (with probability $\rho$, or if at least one
neighbor is in the E state), Q $\rightarrow$ Q (otherwise). To introduce the
so-called ``refractory period'' typical of all excitable systems during which
no re-excitation is possible, the transition from the R state to the Q state is
delayed for $r$ time steps. Thus, the only two parameters in the system are
$\rho$, which  determines the probability that an input to a given site $n$
result into an excitation (i.e., a transition Q $\rightarrow$ E); and $r$
determining the time scale of recovery from the excited state. It turns out
that the precise value of $r$ is not crucial, but choosing a value of $r$ at
least equal or larger than the value of $N$ eliminates a number of numerical
complications \cite{refra}. A dendritic region bombarded by many weak synaptic
inputs is simulated adopting a relatively small value for $\rho$ (here
$10^{-2}$). The typical response of the model under such conditions is
illustrated in Fig. 1. One can see that, starting from arbitrary initial
conditions, eventually an element is first excited (left arrow in Fig. 1). This
initiates a propagated wavefront which collides with others initiated in the
same way somewhere else in the system. After the completion of the refractory
period the process repeats originating another wavefront (right side of Fig.
1). An inmediately apparent feature is the overall similarity
of any two consecutive fronts. The large-scale shape is preserved, despite of
the fact that each element is being randomly perturbed.


We found that important information can be extracted from the analysis of the
dynamics of the first element to be excited in each wavefront, denoted as
$L(n)$. Figure 2 shows  results of numerical simulations where $L(n)$ of each
wavefront is plotted as a function of time. Note the tendency of $L(n)$ to
remain near the previous leading site, which is specially apparent in the
larger systems. To quantify this dynamics we estimated numerically 
$< |L^{t}(n)- L^{t+1}(n)| >$ which is how far (on average) from its current
position the leader will be in the {\sl next} wavefront. The resulting distributions
$<P(\Delta n)>$ of these {\sl jumps} are plotted in Figure 3A for all systems
sizes. The largest probability corresponds to the case in which the wavefront
is first triggered from the same element as in the previous event. The 
power-law $ n^{\pi}$ tells us that there is always a non-zero probability 
for a very long jump, indeed as large as the entire system. Therefore, 
the cut-off of the power law is the only difference between the results obtained 
with small or large N (see Panel A in Figure 3)
Another related measure is the estimation of the average distance the leader
drifts from its current position as a function of time lag $\Delta t$ (t is
always given in  wavefront's units). The results are plotted in panel B of Figure 3. 
The fact that the log-log plot of $|\Delta n| $ vs $\Delta t $ is linear implies  a
power law $\sim \Delta t^{H}$. The best-fit line of the results in Figure 3B 
gives an exponent $H=0.19$. For this case
it is known that the power spectrum decays as $1/f^{\beta}$ and 
that $\beta$ relates with $H$ as $\beta=2H+1=1.4$. 
A random walk will have similar statistical behavior but with an exponent $H=1/2$.
These power-laws, with cutoffs given only by the system size, imply a lack of
characteristic scale (both in time and space), a situation which resembles some
of the scenarios described in the context of self-organized criticality
\cite{soc}.

What causes this memory is trivially simple: the first site to be activated by
the noise will necessarily be the first (exactly after $r$ time steps) to be
recovered and consequently to be ready to be re-excited. The two adjacent sites
which were excited by the leader will recover only after $r+1$ time steps, and
so on for the other adjacent sites. Thus, excitation by the noise will always
be biased by the previous sequence of excitation. Therefore, this ``memory" can
be preserved as long as the cycle of recovery (in this model the $r$ time
steps) is not affected by the noise. Regarding the dependence with the noise
intensity, for vanishingly small $\rho$  all sites will have enough time to
cycle to the Q state and no memory will be kept (see below).
{\sl - Exponents are robust -} The phenomenology described as well as its
characteristics power laws are not model-dependent. The fact  
that similar results were obtained with various numerical models 
motivated the search for the simplest numerical simulation scheme. 
It turns out that the dynamics and the statistical behaviour is preserved 
by the simple kinematic description of the motion 
of these noise-induced propagated excitable waves. The approach 
is described using the cartoon in Figure 4 as follows:
Time and space are considered continuos variables. It is assumed that excitations 
can initiate a wavefront at any point in the 1D space with uniform probability. 
Therefore, after choosing the noise amplitude ($\rho$) the first step of the algorithm is   
to distribute all the potential excitation spots 
at random location and at random times. Larger values of $\rho$ imply more events to be
distributed. Filled circles in Fig. 4 denoted ``a" trough ``e" correspond to few of these events. 
Then the space is scanned to locate the earliest excitation point (i.e., in the figure
is the point ``a" .
Subsequently, two wavefronts are drawn from that point with uniform 
(arbitrary) speed. 
A front dies when either reaches the boundary as in the initial case or upon colliding with 
other front as the one initiated by event labeled ``b". (the dotted lines indicate
two of these interrupted fronts). Thus the algorithm is the repetition of 
scanning follows by the identification of potential collisions. 
It needs to be noted that nothing is peculiar of these rules, simply they are 
the algorithmic description of what is known about excitable waves. 
The results of extensive simulations are plotted in Figure 3 side by side with
those already described for the discrete model. The jump distributions are
plotted for four noise levels in Panel C. The mean drifts  as a function
of time lag are plotted in Panel D. It can be seen that there is a 
remarkable agreement between the numerical values of both scaling exponents.

{\sl -How long does it remember?-} For the sake of demonstration, the dissipation
of memory can be estimated by first imposing an initial activation sequence in
the system (i.e., writing) and then calculating the Hamming distance between
the initial and subsequent wavefront separated by $\Delta t$.
Using the discrete model we impose an arbitrary initial configuration of
excitation, in this case the sinusoidal pattern plotted in the inset of panel
A of Figure 5. As time passes, the pattern deforms as shown by the snapshots at 
times 2, 5, 10 and 50 in the figure, which can be estimated by the Hamming
distance defined as:

\begin{equation} 
< D(t)> = \frac{1}{N}\sum_{n=1}^N
|S^t_n - S^{t+\Delta t}_n| 
\end{equation} 
where $S$ are the initial and subsequent states, ranked by the excitation order
of each element. Means and SEM $D(t)$ were calculated and the results are
plotted in the main body of Figure 5A as a function of $\rho$. It can be seen
that the Hamming distance follows a power-law up to times of about 50 events.
It was already mentioned that for vanishingly small $\rho$ no memory of
previous states can be maintained since this condition implies that all the
element have enough time to go to the Q state preceding the excitation. Thus,
rather paradoxically, more noise implies longer memory. This is illustrated by
the results in Figure 5B, where $D(t)$ was calculated for increasing noise
$\rho$. Thus, we can call this phenomena a form of {\sl noise-induced memory}.

{\sl - Implications for learning and memory -} The  dynamics described here 
might have important consequences for neural ``plasticity''. 
This is the name given, in neuroscience, to the process by which interconnected neurons 
can strengthen or weaken their synaptic contacts to
modulate their communication. The dogma is that memory and learning in animal
brains are based on long-term changes of the synaptic connectivity. An
important point in contemporary thinking assumes that whatever the plastic
process is, it must be able to modify the synaptic strength during the time
window imposed by the longest time-scale in the neuron dynamics. This window
is given by the relaxation kinetics of the membrane and is at most of the order
of hundreds of milliseconds \cite{decay}. This length is considered too short for
producing most of the necessary synaptic changes. 
The length of that window can be many
orders of magnitude longer, if the results presented in this Letter survive the
intricate complexity of the spatial structure in real neurons, as well as 
perhaps other caveats. This would imply that the correlated spatial activity along the
dendrites established by the inputs will only decay after hundreds of firing
events. This time scale is much longer than the fraction of a second currently
considered as the longest time scale that a neuron could remember from its past
history. This correlated sequence of activation can in turn influence the
spatial distribution of the molecular machinery supposedly responsible for the
long-term synaptic modifications.

The work reported here is restricted, for simplicity, to the one dimensional
case and the use of the simplest conceivable excitable model. Nevertheless,
the phenomena is robust and similar results can be obtained using more 
detailed models. If the dynamic described here exist as such in real neurons 
it would be very relevant to neural functioning.

Supported by the Mathers Foundation. R.U. Computer resources are supported by
NSF ARI Grants. Discussions with P. Bak,  R. Llinas and Mark Millonas are appreciated.
Communicated in part by DRC at the {\sl First International Conference on Stochastic
Resonance in Biological Systems. Arcidosso, Italy, May 5-9, 1998} where the
hospitality of the colleagues of the Istituto di Biofisica of Pisa was cherished.

\begin{figure}
\centerline{\psfig{figure=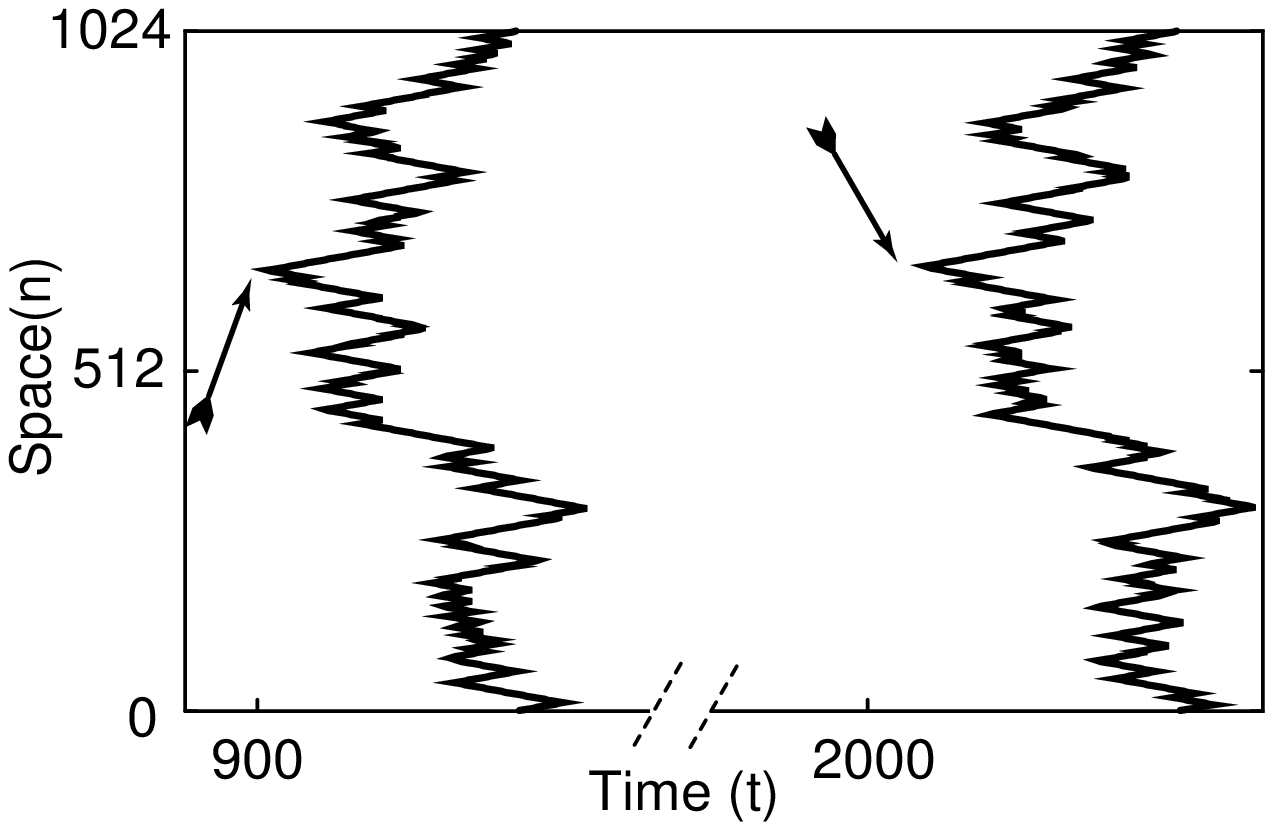,width=3.00 truein}}
\caption{\footnotesize{ An example of two consecutive noise-induced wavefronts.
Note the similarity in the overall shape of the two consecutives wavefronts,
which is typical. The arrows indicate the earliest activated site (i.e., the
leader $L(n)$) in each wavefront.}} 
\end{figure}
\begin{figure}[htbp]
\vspace{-.5in}
\centerline{\psfig{figure=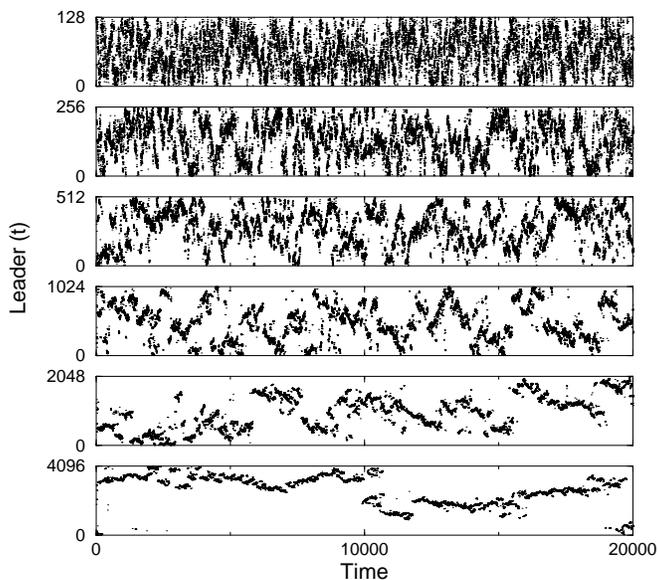,width=3.4in,angle=-90}}
\caption{\footnotesize{Plot of the consecutive positions of the leading element
in each firing event (i.e., the ones identified by the arrows in Fig. 1). The
tendency of the leader is to remain near the previous leading site, a fact
that is visually more apparent in the large systems. (System size increases from
N = 128 at the top to N = 4096 at the bottom panel. $\rho = 10^{-2}$ for all
panels.)}} 
\end{figure} 

\begin{figure}[htbp]
\hspace{-.1in}
\centerline{\psfig{figure=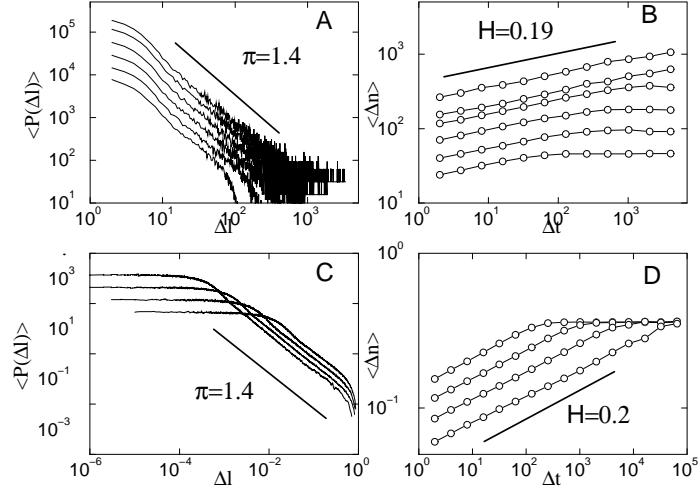,width=3.3in}}
\caption{\footnotesize {(A and C): Distribution of the
differences between $L(n)$ of two consecutive wavefronts.
Results in A correspond to the
discrete model while results plotted in C are from the kinematic
simulation. In both cases the exponent $\pi \sim 1.4$ (B and D): Mean
drift of  $L(n)$
as a function of time lag $\Delta$t. Results in B are from the
discrete model, those plotted in D are from the kinematic simulation.
The mean drift scales as $t^H$, the best-fit line gives $H=0.19$ 
in the case of the discrete model and $H=0.2$ for the results using 
the kinematic description.    
The system sizes for the discrete model  are N = 128, 256, 512, 1024, 2048, 4096 from
bottom to top plots, the noise $\rho=10^{-2}$. For the kinematic 
description system size is fix (unit interval)
and noise density increases
from bottom to top plots from $10^{-6}$; $10^{-5}$ ; $10^{-4}$;$10^{-3}$.
}}
\end{figure}

\begin{figure}[htbp]
\centerline{\psfig{figure=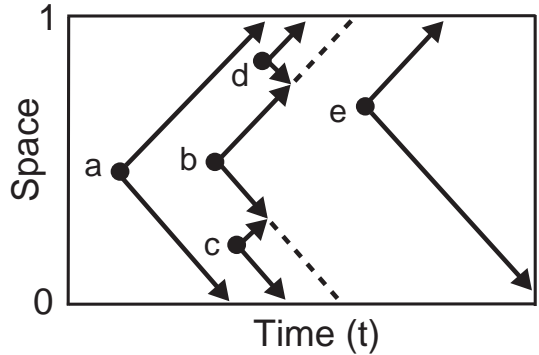,width=2.8in}}
\caption{\footnotesize {Cartoon of the kinematic algorithm (see text).}}
\end{figure}
\begin{figure}[htbp]
\centerline{\psfig{figure=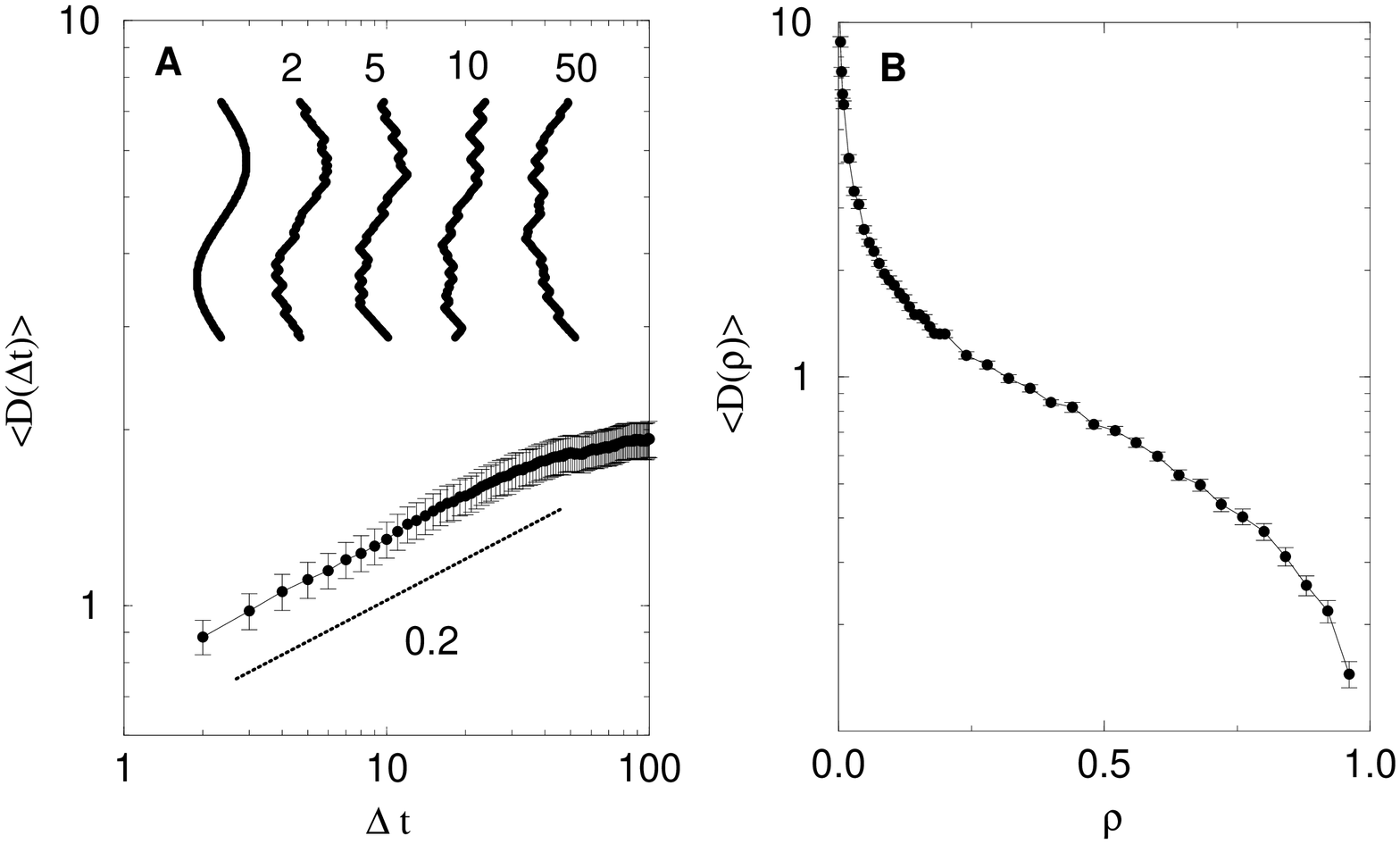,width=2.80truein,angle=-90}}
\caption{\footnotesize{(A): The Hamming distance $< D (t)>$ from
the original sinusoidal pattern as a function of time.
The inset shows the initial wavefront and at time
steps: 2, 5, 10 and 50. $N=256$, $\rho=10^{-2}$ means and SEM of
256 realizations. (B): The Hamming distance between two consecutives
wavefront as a function of noise $\rho$. (Means and SEM of 256 realizations). }}
\end{figure}


\begin{thebibliography}{99}

\bibitem{lapique}H.C. Tuckwell, {\it Stochastic Processes in the
Neurosciences}
(SIAM, Philadelphia, 1989).

\bibitem{active} The reports on the experimental evidence of active
channels
is gargantuan, but a a recent survey can be found in
Z.F.Mainen $\&$ T.J. Sejnowski "Modeling active
dendritic processes in pyramidal neurons." In {\it Methods in Neuronal
 Modeling}. Koch, C, and Segev, I. (eds) 2nd ed., pp.171-210.
MIT Press: Cambridge, MA (1998).

\bibitem{waves} S. Kadar, J. Wang, and K. Showalter. 
Noise Supported Traveling Waves in Subexcitable Media. 
Nature 391,770-772 (1998).

\bibitem{model} J. M. Greenberg and S. P. Hastings.
Spatial patterns for discrete models of diffusion in excitable media.
SIAM J. Appl. Math. {\bf 34}(3),515-523,(1978).

\bibitem{KOCH} C. Koch, {\sl Biophysics of computation}, Oxford
University Press,
New York (1998).

\bibitem{refra}By considering that in these systems the time
to excite the whole system is smaller than the recovery time we
could safely assume  $r > N$.

\bibitem{othermodels}Other numerical
formulations including a explicit integrate and fire scheme give
similar results.


\bibitem{soc} P.  Bak, C.  Tang, and K.  Wiesenfeld,
Phys.  Rev. Lett. {\bf 59}, 381 (1987);
Phys. Rev. {\bf A38}, 364 (1988);
P. Bak, {\it How nature works: the science of
self-organised criticality.} (Springer, New York, 1996;
Oxford University Press, 1997);  M. Paczuski, S. Maslov, and P. Bak,
Phys. Rev. {\bf E53}, 414 (1996).

\bibitem{Mel} B.W. Mel, J. Neurophys. {\bf 70},1086-1101 (1993)

\bibitem{decay}Basically one have to consider the
decay to equilibrium of the membrane potential after being perturbed by
the synaptic current. A rough figure of hundreds of milliseconds is the
current estimate.

\end{thebibliography}
\end{document}